\documentclass{elsart}
\usepackage[]{epsfig}
\begin{document}

\begin{frontmatter}

\title{Un-graviton corrections to the Schwarzschild black hole}

\author{Patricio Gaete\thanksref{cile}}
\thanks[cile]{e-mail address: patricio.gaete@usm.cl}
\address{Departmento de F\'{\i}sica and Centro
    Cient\'{\i}fico-Tecnol\'{o}gico
    de Valpara\'{\i}so, Universidad T\'ecnica Federico Santa
    Mar\'{\i}a, Valpara\'{\i}so, Chile}

\author{Jos\'{e} A. Hela\"{y}el-Neto\thanksref{cbpf}}
\thanks[cbpf]{e-mail address: helayel@cbpf.br }
\address{Centro Brasileiro de Pesquisas
F\'{\i}sicas, Rua Xavier Sigaud, 150, Urca, 22290-180, Rio de
Janeiro, Brazil}

\author{Euro Spallucci\thanksref{infn}}
\thanks[infn]{e-mail address: spallucci@ts.infn.it }
\address{Dipartimento di Fisica Teorica, Universit\`a di Trieste
and INFN, Sezione di Trieste, Italy}

\begin{abstract}
We introduce an effective action  smoothly extending
the standard Einstein-Hilbert action to include  \emph{un-gravity}
effects. The improved field equations are solved for the \emph{Un-graviton corrected}
Schwarzschild geometry reproducing the Mureika result. This is an important
test to confirm the original ``guess'' of the form of the Un-Schwarzschild metric.
Instead of working in the weak field approximation and ``dressing'' the Newtonian
potential with un-gravitons, we solve the ``effective Einstein equations'' including
all order un-gravity effects. An unexpected ``bonus'' of accounting un-gravity 
effects is the \textit{fractalisation} of the event horizon. 
In the un-gravity dominated regime the event horizon thermodynamically
behaves as fractal surface of dimensionality twice the scale dimension $d_U$.

\end{abstract}
\end{frontmatter}

\section{Introduction}
Scale invariance plays an important role in different sectors of 
modern theoretical physics. In statistical mechanics
fluctuations occur at all scale near a critical point and
phase transitions require a scale invariant description.\\
In field theory scale invariance is related to the behavior
of the theory under dilatations and indicates the absence of
a fundamental length scale. \\
For later convenience, it can be useful to recall that in Mathematics
scale invariance is tightly linked to the self-similarity
of fractal curves and surfaces.\\
Finally, scale invariant quantum field theories generally describes
massless particles. Only, recently a new implementation of
this symmetry has been suggested \cite{Georgi:2007ek,Georgi:2007si}
as a consequence of non-trivial fixed point in the infrared
regime \cite{Banks:1981nn}. In this new framework 
 the intuitive notion of scale invariance as a property of massless particles only, 
 has been extended to a new kind of stuff  with no definite mass at all. For this reason, 
 this presently unknown, scale invariant, sector of the elementary
particle spectrum has been dubbed as the \textit{un-particle} sector. \\
The proposal is that below
some critical energy scale $\Lambda_U$ the standard model particles
can interact with un-particles. The forthcoming start of LHC activity
has focused the interest of the high energy physics
community on possible experimental signature of un-particles events
at the $TeV$ energy scale
\cite{Strassler:2008bv,Liao:2007ic,Liao:2007fv,Liao:2007bx,Rizzo:2007xr}
\cite{Cheung:2007ap,Bander:2007nd,Cheung:2007jb,Cheung:2007zza}
\cite{Kikuchi:2007qd,Kikuchi:2008pr}.
Astroparticle and cosmological un-particle effects have been considered as well
\cite{Kikuchi:2007az,Lewis:2007ss,McDonald:2007bt,Davoudiasl:2007jr},
\cite{Das:2007nu,Alberghi:2007vc,Hannestad:2007ys,Freitas:2007ip,Chen:2007qc},
\cite{Bertolami:2009qn,Das:2007cc}.\\
On the  theoretical side,  interesting connections
between un-particles and non-standard Kaluza-Klein dynamics in
extra-dimensions, and AdS/CFT duality, have been pointed out in
\cite{Stephanov:2007ry,Kazakov:2007fn,Kazakov:2007su,Lee:2007xd,Gaete:2008aj}.
 Even deeper connections between un-gravity and trans-Planckian physics
 are currently under investigation \cite{Nicolini:2010bj,Nicolini:2010nb}.
 Finally, two of us (P.G. and E.S.) have built an ``effective action''
 for several kind of un-particle fields, including un-gravitons \cite{Gaete:2008wg}  .
 Encoding un-gravity into an effective action can be useful in view of
studying gravitational effects \cite{Goldberg:2007tt,Mureika:2007nc}
beyond the weak field approximation.\\
In this letter we show that the un-Schwarzschild metric guessed in \cite{Mureika:2007nc}
through perturbative arguments is an \textit{exact} solution of the field equations
obtained from the effective theory introduced in \cite{Gaete:2008wg}. A non-trivial
step in obtaining the solution is the introduction of a point-like source in the
un-gravity Einstein equations. \\
In the final part of this communication we recover ``from scratch'' the area law for the
un-Schwarzschild black hole. The resulting expression for the area suggests the
horizon to be a \textit{fractal} surface of dimension twice the scale dimension $d_U$.

\section{Un-gravity field equations}

The physical system we are going to investigate is an ``hybrid'' of classical matter, 
classical gravity,
and ``quantum'' un-gravity due to the exchange of un-gravitons. The action for this system 
is the sum
of a classical functional $S_M$ for matter, and a \textit{non-local} effective action $S_U$ 
smoothly
extending the Einstein-Hilbert action to include un-gravitons dynamics.
\begin{equation}
 S\equiv S_M + S_U
\end{equation}
$S_M$ is the classical matter action for a massive, point-like, particle ``sitting'' in the 
origin. There
is some freedom to choose the  explicit for of this functional. Simplicity suggests to 
introduce $S_M$
in the form of the action for pressure-less, static fluid,  with a ``singular'' 
( but integrable ! ) energy density mimicking a ``point-mass'':
\begin{equation}
 S_M\equiv -\int d^4x \sqrt{g}\,\rho\left(\, x\,\right)\, u^\mu\, u^\nu\ ,\quad 
 \rho\left(\, x\,\right)\equiv
 \frac{M}{\sqrt{g}}\int d\tau \,\delta\left(\, x -x\left(\tau\right)\,\right)
\end{equation}

The un-gravity action is obtained by combining the Einstein-Hilbert functional and the 
non-local effective action we obtained in \cite{Gaete:2008wg} :
\begin{equation}
S_U = \frac{1}{2\kappa^2}\,\int d^4x \sqrt{g}\,\left[ \,
1+\frac{ A_{d_U}}{\left(\,2d_{U}-1\,\right)\sin\left(\,\pi\, d_U\,\right)}
\frac{\kappa_\ast^2}{\kappa^2}
\left(\, \frac{-D^2}{\Lambda^2_U}\,\right)^{1-d_U}\, \right]^{-1} R
\label{ueh}
\end{equation}

where, $D^2$ is the generally covariant D'Alembertian;

\begin{equation}
A_{d_U}\equiv \frac{16\pi^{5/2}}{\left(\, 2\pi\,\right)^{2d_U}}
\frac{\Gamma\left(\,d_U + 1/2\,\right)}{ \Gamma\left(\,d_U - 1\,\right)
\Gamma\left(\,2d_U \,\right)}
 \end{equation}

while $\kappa_\ast$ represents
the  un-gravitational Newton constant  

\begin{eqnarray}
\kappa_\ast &&\equiv \frac{1}{\Lambda_U}\left(\,
\frac{\Lambda_U}{M_U}\,\right)^{d_{UV}}\\
 &&\simeq \frac{1}{\Lambda_U}\left(\,
\frac{\Lambda_U}{M_U}\,\right)\ ,\qquad d_{UV}\simeq 1
\end{eqnarray}

The strength of coupling constant is determined by the mass scale $M_U$
which replaces the Planck mass. $\kappa=16\pi\, G_N=M_{Pl.}^{-2}$.

 Our main
purpose is to solve the field equations derived from $S$ by assuming
the source is static, i.e. the four-velocity field $u^\mu$ has only 
non-vanishing time-like component

\begin{equation}
u^\mu\equiv \left(\, u^0\ , \vec{0} \,\right)\ ,\quad
u^0=\frac{1}{\sqrt{-g^{00}}}
\end{equation}

Einstein equations are obtained by varying the action (\ref{ueh})
with respect to the metric $g_{\mu\nu}$. By neglecting surface terms
coming from the variation of the generally covariant D'Alembertian, we find

\begin{eqnarray}
 R^\mu_\nu-\frac{1}{2}\delta^\mu_\nu\, R &&= \kappa^2\,
 \left[ \, 1+
 \frac{A_{d_U}\Lambda^{2-2d_U}_U}{\left(\, 2d_U-1\,\right)\,\sin\left(\,\pi\, d_U\,\right) }
  \frac{\kappa_\ast^2 }{\kappa^2}
\left(\,-D\,\right)^{d_U-1}\,\right] \,
T^\mu{}_\nu\nonumber\\
&&\equiv \kappa^2\, T^\mu{}_\nu +\kappa_\ast^2 \frac{A_{d_U}}{ \sin\left(\,\pi\, d_U\,\right) }
T_U{}^\mu_\nu
\label{e1}
\end{eqnarray}

In Eq.(\ref{e1}) we have ``shifted'' the un-particle terms to the 
 r.h.s. leaving the l.h.s. in the canonical form. As a matter of fact,
Eq. (\ref{e1}) can be seen as `ordinary'' gravity coupled to an
``exotic'' source term, instead of un-gravity produced by an ordinary
particle. The two interpretations are physically equivalent.\\
Now, the reader is probably thinking ``... but, in the
Einstein field equations for the Schwarzschild solution there is
no energy-momentum tensor. Schwarzschild geometry
is a \textit{vacuum} solution.''  It is a quite common misunderstanding
to think   that vacuum means ``in the absence of a source''
instead of  \textit{outside} a compact (localized)  source. 
In the Schwarzschild case,
the source is a point-like particle sitting in the origin
\footnote{In General Relativity textbooks it is customary to introduce
the Schwarzschild solution without even mentioning the presence of
a point-like source. Once the Einstein equations are solved in the vacuum,
the integration constant is determined by matching the solution with
the Newtonian field outside a spherically symmetric mass distribution.
Definitely, this is not the most straightforward way to expose
students, and not only them, to one of the most fundamental solutions
of the Einstein equations. Moreover, the presence of
a curvature singularity in the origin, where from the very
beginning a finite mass-energy is squeezed into a zero-volume point,  
is introduced as a ``shocking'',  un-expected result. Against this
background, we showed in 
\cite{Nicolini:2005vd,Spallucci:2006zj,Ansoldi:2006vg,Spallucci:2008ez,Nicolini:2009gw} 
that once quantum delocalization of the source is accounted, all these ``flaws'' disappear.}.
The corresponding energy-momentum tensor is given by \cite{DeBenedictis:2007bm}

\begin{eqnarray}
&& T^0_0=-\frac{M}{4\pi\, r^2}\delta\left(\, r\,\right) \label{t00}\\
&& T^r_r=0\\
&& T^\theta_\theta=T^\phi_\phi= -\frac{M}{16\pi\, r}\delta\left(\, r\,\right)
\frac{1}{g_{00}}\partial_r\, g_{00}
\end{eqnarray}

where, $T^\theta_\theta\ ,T^\phi_\phi$ are determined by the requirement
$\Delta_\mu T^{\mu\nu}=0$.\\
With this kind of energy-momentum tensor the $00$ and $rr$ components
of the metric tensor turn out to be of the form
\begin{equation}
g_{rr}^{-1}= 1- \frac{2G_N}{r}\, M\left(\, r\,\right)=-\frac{e^{-h_0}}{g_{00}} 
\end{equation}

where the constant $h_0$ can be freely re-absorbed into the definition of
the time coordinate, and

\begin{equation}
M\left(\, r\,\right)\equiv 4\pi\,\int dr\, r^2 \, T^0_0 \label{linm}
\end{equation}

In Equation (\ref{linm}) the symbol $\int dr$ indicates an indefinite 
integration. 
The constant factor $e^{h_0}$ can be safely rescaled to $1$ by
a redefinition of the time coordinate.\\

As a first step towards solving the Einstein field equation 
we need to transform the energy density (\ref{t00}) into its
un-particle counterpart $T^0_{U\, 0}$. We momentarily switch to
isotropic, (free-falling) Cartesian-like coordinate for computational 
convenience.
Finally, it will be easy to transform the result in a spherical frame. 
By taking into account (\ref{e1}) and(\ref{t00}) we can write

\begin{eqnarray}
T^0_{U\, 0} &&\equiv \rho_U= \frac{M}{2d_U-1}\,\Lambda^{2-2d_U}_U 
\left(-\nabla^2 \,\right)^{d_U-1}\,\delta\left(\,\vec{x}\,\right)
\\
&&=\frac{M}{2d_U-1}\,\frac{\Lambda^{2-2d_U}_U}{\left(\,2\pi\right)^3}\int d^3k\,\left(\,
\frac{1}{\vec{k}^2}\,\right)^{1-d_U}\, e^{i\vec{k}\cdot\vec{x}}
\label{t002}
\end{eqnarray}

In order to proceed we use the Schwinger representation for 
$\left(\,1/{\vec{k}}^{\, 2}\,\right)^{1-d_U}$:

\begin{equation}
\left(\,\frac{1}{{\vec{k}}^{\, 2}}\,\right)^{1-d_U}=
\frac{1}{\Gamma\left(\,1-d_U \,\right)}\int_0^\infty ds s^{-d_U}\, 
e^{-s\,\vec{k}^{\,2}}\label{schw}
\end{equation}

By inserting (\ref{schw}) in (\ref{t002}), integrating first over
$\vec{k}$ and secondly over the Schwinger parameter, we get $\rho_U $

\begin{equation}
\rho_U\left(\,\vec{x}\,\right)=
\frac{2^{2d_U}}{16\pi^{3/2}} 
\frac{\Gamma\left(\,d_U +1/2\,\right)}{\Gamma\left(\,1-d_U \,\right) }
\, \frac{M}{2d_U-1} \Lambda^{2-2d_U}_U \left(\, \frac{1}{\vec{x}^{\,2}} \,\right)^{d_U+1/2}
\end{equation}

which is manifestly spherically symmetric and can be conveniently
written in terms of the radial distance from the origin $r\equiv
\vert\vec{x}\vert $:

\begin{equation}
\rho_U\left(\,r\,\right)=
\frac{2^{2d_U-1}}{16\pi^{3/2}} 
\frac{\Gamma\left(\,d_U -1/2\,\right)}{\Gamma\left(\,1-d_U \,\right) }
\,M\, \Lambda^{2-2d_U}_U \left(\,\frac{1}{r} \,\right)^{2d_U+1}
\end{equation}

Thus,

\begin{eqnarray}
M\left(\, r\,\right)&&=4\pi\,\frac{2^{2d_U-1}}{16\pi^{3/2}} 
\frac{\Gamma\left(\,d_U -1/2\,\right)}{\Gamma\left(\,1-d_U \,\right) }
\, M \Lambda^{2-2d_U}_U\, \int dr \, r^{1-2d_U}\nonumber\\
&&=\frac{2^{2d_U-1}}{4\pi^{1/2}} 
\frac{\Gamma\left(\,d_U -1/2\,\right)}{\Gamma\left(\,1-d_U \,\right) }
\, M \Lambda^{2-2d_U}_U\,\frac{r^{2-2d_U}}{2\left(\, 1-d_U\,\right)}
\end{eqnarray}

\begin{figure}[h]
\begin{center}
\includegraphics[width=10cm,angle=0]{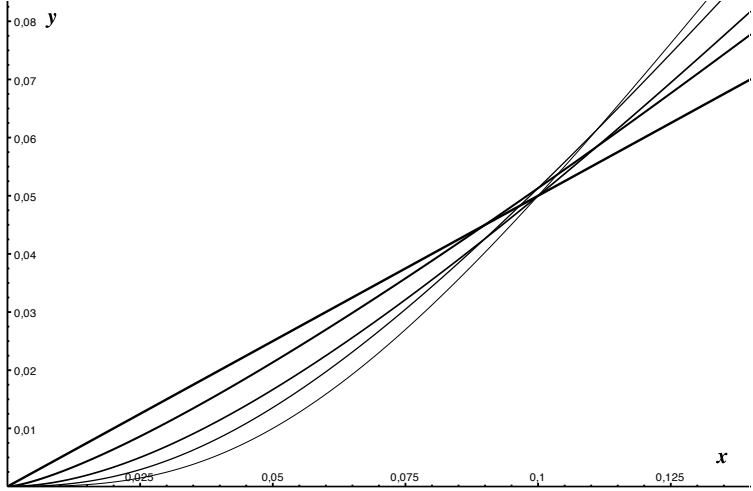}
\caption{\label{horizon} 
\textit{ Plot of the function $M(r_H)$ for different values of $d_U$. The straight line 
corresponds to
$d_U=1$, the other lines have $d_U=1.25\ , 1.5\ , 1.75\ , 2 $.  } }
\end{center}
\end{figure}

Finally,

\begin{equation}
M\left(\, r\,\right)=
\frac{2^{2d_U-2}}{4\pi^{1/2}} 
\frac{\Gamma\left(\,d_U -1/2\,\right)}{\Gamma\left(\,2-d_U \,\right) }
\, M \Lambda^{2-2d_U}_U\,\left(\,\frac{1}{r}\,\right)^{2d_U-2}
\end{equation}

An arbitrary integration constant has been set to zero, as we are
interested to study the field determined in a unique way by an un-particle
source, and nothing else.\\
Thus, we find

\begin{eqnarray}
g_{rr}^{-1}&&= -g_{00}= 1 -\frac{2M G_N}{r}\left[\, 1 + \kappa_*^2\,
\frac{A_{d_U}}{\sin\left(\,\pi\, d_U\,\right)}
\frac{ M\left(\, r\,\right)}{2M\, G_N}\,\right]\nonumber\\
&&=1- \frac{R_s}{r}\left[\, 1+\frac{M^2_{Pl.}\kappa_\ast^2}{\pi^{2d_U-1}}
\,\Lambda^{2-2d_U}_U \frac{\Gamma\left(\,d_U -1/2\,\right)\Gamma\left(\,d_U +1/2\,\right)}
{\Gamma\left(\,2d_U\,\right)}
\,\left(\,\frac{1}{r}\,\right)^{2d_U-2}\,\right]\nonumber\\
&&=1- V_N\left(\, r\,\right)\left[\, 1+ \Gamma_U\,\left(\,\frac{R_\ast}{r}\,\right)^{2d_U-2}
\,\right]
\label{unschw}\\
\Gamma_U &&\equiv \frac{2}{\pi^{2d_U-1}}
\frac{\Gamma\left(\,d_U -1/2\,\right)\Gamma\left(\,d_U +1/2\,\right)}
{\Gamma\left(\,2d_U\,\right)}
\end{eqnarray}

where, $R_s=2MG_N=2M/M^2_{Pl.}$ is the Schwarzschild radius;
$-V_N\left(\, r\,\right)$ is the Newton gravitational potential, and
the new gravitational length scale $R_\ast$ is defined as

\begin{equation}
R_\ast\equiv \frac{1}{\Lambda_U}\left(\,\frac{M_{Pl.}}{M_U}\,\right)^{1/(d_U-1)}
\end{equation}
The metric (\ref{unschw}) has been guessed in the weak-field case \cite{Mureika:2007nc} from 
the form of the un-graviton dressed Newtonian potential. At the linearized level
the correction to the Newton law is of the form

\begin{equation}
V_U\left(\, r\,\right)=-G_U\frac{m_1 m_2}{r^{2d_U-1}}
\end{equation}

In ref.\cite{Das:2007cc} the phenomenological consequences of $V_U(r)$ have been discussed
and the correction to the perihelion precession of Mercury has been obtained

\begin{equation}
\delta \theta\simeq 2\pi\, \left( d_U-1\,\right)\left( 2d_U-1\,\right) \frac{V_U}{V_N}
\end{equation}

The same result can be obtained from the exact solution, which is
valid for any strength of the gravitational field, we obtained in this paper.\\
The horizon curve is obtained by the condition $g_{rr}^{-1}(r_H)=0$ 

\begin{equation}
 M=\frac{r_H}{2G_N}\frac{1}{1 + \Gamma_U\,\left(\,R_\ast/r_H\,\right)^{2d_U-2}}
\end{equation}

For $d_U=1$ the horizon radius result to be increased with respect to the pure-gravity case 
as one finds

\begin{equation}
 r_H= R_s \left(\, 1 + \Gamma_1\,\right)
\end{equation}

\begin{figure}[h]
\begin{center}
\includegraphics[width=10cm,angle=0]{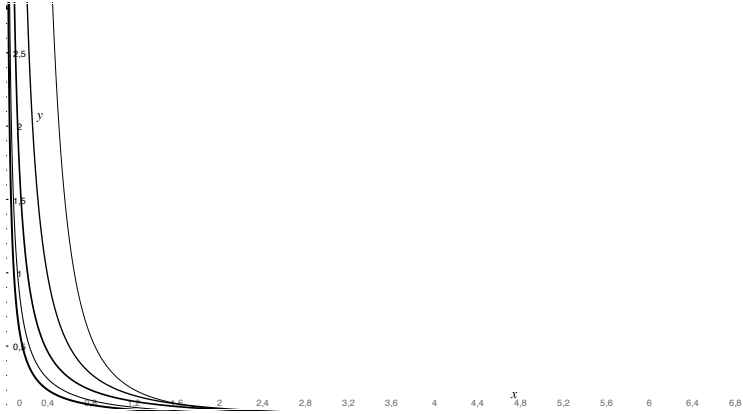}
\caption{\label{Thorizon} 
\textit{Plot of the function $T_{d_U}(r_H)$ for different values of 
$d_U=1\ ,1.25\ , 1.5\ , 1.75\ , 2 $.}} 
\end{center}
\end{figure}

The Hawking temperature is

\begin{eqnarray}
 T_{d_U}&&=\frac{1}{4\pi r_H}\frac{1}{\left[\, 1 + \Gamma_U 
 \left(\, R_\ast/r_H\,\right)^{2d_U-2}\,\right] }
 \left[\, 1 + \left(\, 2d_U-1\, \right)\, \Gamma_U\,  
 \left(\, \frac{R_\ast}{r_H}\,\right)^{2d_U-2} \,\right]\nonumber\\
 &&=\frac{1}{4\pi r_H}\left[\, 
 1 + 
\frac{2\left(\, 2d_u-1\,\right) \,\Gamma_U}{ 1+ \Gamma_U\,\left(\, R_\ast/r_H \,\right)^{2d_U-2} }
 \,\left(\, \frac{R_\ast}{r_H}\,\right)^{2d_U-2} \, \right]
\end{eqnarray}

We can distinguish two different ``phases'' of the model :\\
i) \textit{gravity-dominated} phase, where $T_{d_U}$ takes the standard form

\begin{equation}
 T_{d_U}\simeq T_H=\frac{1}{4\pi r_H}
\end{equation}
 
ii ) \textit{un-gravity-dominated} phase, where $T_{d_U}$ turns into

\begin{equation}
 T_{d_U}\simeq \frac{2d_U-1}{4\pi r_H} \label{unt}
\end{equation}

It is interesting to compare Eq.(\ref{unt}) with the corresponding temperature for a
Schwarzschild black hole in $D$ spacetime dimensions

\begin{equation}
 T_{D}\simeq \frac{D-3}{4\pi r_H} \label{td}
\end{equation}

More precisely, Eq.(\ref{td}) gives the intrinsic temperature of a $D-2$ dimensional horizon.
By comparing Eq.(\ref{unt}) with Eq.(\ref{td}) we see that in the ungravity dominated phase
the thermodynamic behavior of the horizon corresponds to a an \textit{effective, non-integer
dimension} $d_H=2d_U$. Under this respect, ungravity leads to a \textit{fractalisation} of the
event horizon. Let us elaborate this picture by investigating the Area Law.\\
It is ``customary'' nowadays to assume that the black hole entropy is $1/4$ of the event 
horizon
area in Planck units, forgetting that this is a \textit{consequence} of the First Law
of black hole thermodynamics and not a ``dogma'' to be blindly accepted.\\
Against this background, we start from

\begin{equation}
 dM = T_{d_U} dS
\end{equation}
 
and derive $S$ for a ungravity dominated black hole. In this regime we can express $M$ in 
terms of $r_H$ as

\begin{equation}
 M\simeq \frac{R_\ast}{2\Gamma_U} \left(\,r_H/ R_\ast \,\right)^{2d_U-1}
\end{equation}

The First Law takes the form 
 
\begin{equation}
dS= \frac{dM}{T_{d_U}} =\frac{4\pi\, r_H}{2d_U -1}\, \frac{\partial M}{\partial r_H}\, dr_H
\label{ds}
\end{equation}
 
By integrating (\ref{ds}) we find

\begin{equation}
 S= \frac{\pi\, R_\ast^{2-2d_U}}{d_U\,\Gamma_U}\, r_H^{d_H}
 \label{unlaw}
\end{equation}

Eq.(\ref{unlaw}) represents the Area Law for an Un-Schwarzschild black hole. It is immediate
to check that for $d_U\to 1$ we recover the standard form

\begin{equation}
 S\to \pi\, r_H^2 =\frac{1}{4} A_H
 \label{1/4}
\end{equation}

Thus, we feel confident to interpret Eq.(\ref{unlaw}) as the extension of the Area Law for a
fracatlised horizon of dimension $d_H=2d_U$.\\
The analogy between tensor un-particle dynamics and physics in presence of fractal
extra-dimension has been noted in \cite{Mureika:2008dx} in relation with multiplicity,
temperature profile and decay rate of $TeV$ micro black holes. \\
Our result trace back these effects not to a formal analogy with black holes in higher 
dimensional
spacetime, but  to the fractal geometry of the event horizon itself. The out-coming picture
is that if we imagine the event horizon as a null surface ``built-up'' of un-gravitons trapped at the
Schwarzschild radius, the underlying scale invariance of the theory manifests itself in the 
form of fractality, or self-similarity, of the horizon.\\
The presence of a source in the effective Einstein equations is instrumental to evaluate
un-gravity corrections. In this paper we considered the simplest case of a point-like
source leading to an ``eternal'' black hole type solution. An interesting problem
is the study a collapsing body in order to see if and how un-gravity can change the
dynamics of the collapse itself. Even in the simplest case of a 
collapsing shell of matter, it is  non-trivial to extend the Israel matching
formalism to the case of our effective, non-local, theory. Thus, we postpone this
study to a future investigation.
A further prospect of pushing forward the study of un-particle fields 
in connection with gravity, we believe it would
be interesting to build up an un-gravity action taking into account
new gravitational degrees of freedom which drop out whenever we
allow for dynamical torsion. This would lead to a richer spectrum and
the un-particle sector might present some peculiar properties.

\end{document}